# Phantom-based gradient waveform measurements with compensated variable-prephasing: Description and application to EPI at 7T

**Hannah Scholten[1], Tobias Wech[1,2], Istvan Homolya[3], and Herbert Köstler[1]**

[1]Department of Diagnostic and Interventional Radiology, University Hospital Würzburg, Würzburg, Germany

[2]Comprehensive Heart Failure Center, University Hospital Würzburg, Würzburg, Germany

[3]Chair of Molecular and Cellular Imaging, Comprehensive Heart Failure Center, University Hospital Würzburg, Würzburg, Germany

**Correspondence:**

Hannah Scholten

Department of Diagnostic and Interventional Radiology

University Hospital Würzburg

Oberdürrbacher Str. 6, Würzburg, 97080, Germany

E-Mail: scholten_h@ukw.de

**Abstract**

**Purpose:** Introducing "compensated variable-prephasing" (CVP), a phantom-based method for gradient waveform measurements. The technique is based on the "variable-prephasing" (VP) method, but takes into account the effects of all gradients involved in the measurement.

**Methods:** We conducted measurements of a trapezoidal test gradient, and of an EPI readout gradient train with three approaches: VP, CVP, and "fully compensated variable-prephasing" (FCVP). We compared them to one another and to predictions based on the gradient system transfer function. Furthermore, we used the measured and predicted EPI gradients for trajectory corrections in phantom images on a 7T scanner.

**Results:** The VP gradient measurements are confounded by lingering oscillations of the prephasing gradients, which are compensated in the CVP and FCVP measurements. FCVP is vulnerable to a sign asymmetry in the gradient chain. However, the trajectories determined by all three methods resulted in comparably high EPI image quality.

**Conclusion:** We present a new approach allowing for phantom-based gradient waveform measurements with high precision, which can be useful for trajectory corrections in non-Cartesian or single-shot imaging techniques. In our experimental setup, the proposed "compensated variable-prephasing" method provided the most reliable gradient measurements of the different techniques we compared.

**Keywords:** gradient measurement, ultrahigh field, variable-prephasing, gradient impulse response, EPI

**Introduction**

The dynamically switching gradient fields of an MRI scanner can suffer from temporal errors due to hardware delays, eddy currents, coil vibrations, or other system imperfections. Uncorrected, erroneous gradient waveforms can cause image artifacts, especially in non-Cartesian or single-shot techniques. Therefore, methods to obtain precise knowledge of the actual gradient field evolution are becoming increasingly important and popular.

Field cameras are one possibility to measure the true gradient waveforms in the bore of the scanner.[1,2] However, they require special hardware and software, which are not available at every MR site. A very simple gradient measurement method, which only uses standard scanner hardware, is the so-called thin-slice method, where the gradient waveform is inferred from the change in the signal phase of a simple FID measurement in a thin off-center slice.[3] However, this approach is limited by the gradient-induced signal dephasing in the excited slice.[4] When the signal magnitude becomes too small, the phase cannot be determined correctly anymore, which ultimately leads to erroneous gradient waveform estimates. Another gradient measurement principle is summarized under the term self-encoding methods.[4–8] They apply a dephasing gradient with a defined integral between a slice-selective excitation and the FID readout during which the gradient of interest is applied. By repeating this with a sufficient number of different amplitudes of the self-encoding gradient, the actually applied gradient waveform can be inferred from the envelope of the magnitude signals. The major drawback of this approach is the long acquisition time, because it typically requires a large number of repetitions. Recently, Harkins and Does published a hybrid method combining the thin-slice method with the self-encoding approach. By applying self-encoding gradients of variable amplitudes in repeated acquisitions of the thin-slice signal, their so-called variable-prephasing approach enables gradient waveform measurements in thicker slices with higher SNR than in the thin-slice method, while keeping the scan time relatively short. However, they neglect the effects of the "prephasing" gradients, which can lead to inaccurate results.[9]

In this study, we present "compensated variable-prephasing" (CVP), an extension of the variable-prephasing method[4] for phantom-based measurements of arbitrary gradient waveforms. Our approach allows for the compensation of effects that are neglected in the original method, namely lingering repercussions of the prephasing gradients. Another method that we developed in a previous study[10] and term "fully compensated variable-prephasing" (FCVP) additionally compensates for concomitant field effects. We evaluate the differences between the three methods, and compare their results to the predictions of a linear time-invariant model of the gradient system.[11] This model is represented by the gradient system transfer function (GSTF).[12–14] We also assess how the differences between the presented methods affect actual image reconstructions. We therefore conducted EPI experiments in a phantom on a 7T scanner and applied trajectory corrections based on these methods. Parts of this work have previously been presented at the Annual Meeting of the ISMRM 2023.[9]

**Theory**

The variable-prephasing method[4] constitutes an extension of the thin-slice method[3,15] and shall be briefly reviewed here before introducing the "compensated" and "fully compensated variable-prephasing" approaches. In the thin-slice method, slice-selective excitations are followed by FID readouts during which the test gradient waveform of interest is played out. The gradient waveform is then determined from the phase difference of the signals in the excited slices. Depending on the applied gradient moment and the slice thickness, this can eventually lead to complete signal dephasing, which prevents a meaningful phase extraction from the signal. The variable-prephasing approach lifts this limitation by adding prephasing gradients of variable amplitude between the excitation and the test gradient. Similar to self-encoding methods for gradient waveform measurements[5–8], this technique creates a maximum in the acquired FID signal when the integral of the test gradient waveform is equal to the negative integral of the prephasing gradient. Repeating the acquisition with different prephasing amplitudes results in maxima at different time points during the test gradient and allows to compensate for the dephasing effect of the test gradient itself. The acquisition scheme is depicted in boxes (1) and (2) of Figure 1A. A reference acquisition without the prephasing and test gradients is used to account for effects from the slice selection gradient (box 2). Details on how to determine the waveform can be found in the original publication.[4]

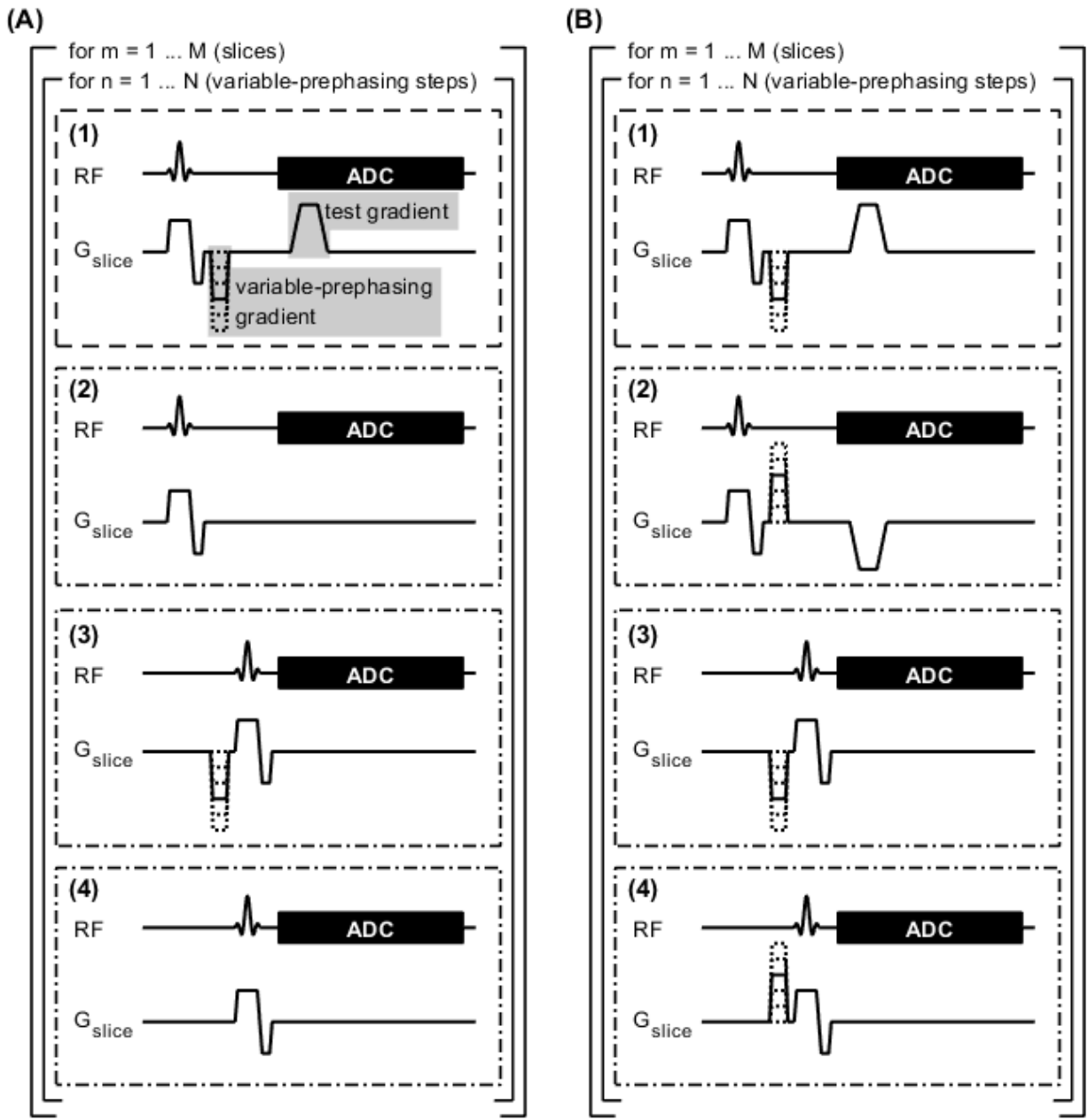

Figure 1. Sequence diagrams demonstrating the measurement schemes of our (A) "compensated variable-prephasing" (CVP) and (B) "fully compensated variable-prephasing" (FCVP) approaches. CVP accounts for lingering effects of the slice selection and the prephasing gradients. FCVP additionally compensates concomitant field effects.

The variable-prephasing method accounts for lingering effects of the slice selection gradient. However, it neglects effects from the prephasing gradients. Addressing them is the main purpose of the proposed "compensated variable-prephasing" (CVP) approach. To this end, we added two more steps to the measurement scheme, which are depicted in boxes (3) and (4) of Figure 1A. In step (3), we apply the prephasing gradient before the slice selective excitation, which allows us to capture lingering effects from the prephasers at the time of the FID readout.[10,16] Finally, to account for the effects of the slice selection gradient in step (3), we repeat the acquisition without the prephasing gradient in step (4). It is important that the timing of the prephasing gradients and the readout does not change, so in step (1), there is a delay between the prephasing gradient and the readout, long enough to be filled by the slice selection gradient and its rephaser in step (3).

To transfer into "fully compensated variable-prephasing" (FCVP), we modified steps (2) and (4) in a way that allows removing concomitant field effects from the final waveform, similar to the method published by Brodsky et al.[15] Instead of conducting the reference measurements with just the slice selection active, we invert the sign of the prephasing and test gradients in the reference measurements. Since concomitant gradient fields depend approximately quadratically on the gradient amplitude[17], they can be cancelled out by subtracting the signals from acquisitions with inverted gradients. The measurement scheme for FCVP is shown in Figure 1B. In the following, we develop the mathematical description of FCVP, and successively reduce it to obtain the formulas for CVP and VP.

To determine the test gradient's field evolution, we consider the phase of the measured FID signal and take the time derivative. The resulting frequency evolution, denoted $f(r,t)$, is described by the following equations in the four measurement steps of the n$^{th}$ prephasing step depicted in Figure 1B:

$$f_{n,1}(r,t) = \frac{\gamma}{2\pi}\left[\sum_{j=1}^{2} p_j(r)(d_j(t) + d_{j,n}^{VP}(t))\right] + c(r,t) + c_n^{VP}(r,t) + q^I(r,t) \tag{1}$$

with $p_1(r) = 1, p_2(r) = r, d_1(t) = \Delta B_0(t), d_2(t) = G(t), d_{1,n}^{VP}(t) = \Delta B_{0,n}^{VP}(t), d_{2,n}^{VP}(t) = G_n^{VP}(t)$

$$f_{n,2}(r,t) = \frac{\gamma}{2\pi}\left[\sum_{j=1}^{2} p_j(r)(-d_j(t) - d_{j,n}^{VP}(t))\right] + c(r,t) + c_n^{VP}(r,t) + q^I(r,t) \tag{2}$$

$$f_{n,3}(r,t) = \frac{\gamma}{2\pi}\left[\sum_{j=1}^{2} p_j(r)d_{j,n}^{VP}(t)\right] + c_n^{VP}(r,t) + q^{II}(r,t) \tag{3}$$

$$f_{n,4}(r,t) = \frac{\gamma}{2\pi}\left[\sum_{j=1}^{2} p_j(r)(-d_{j,n}^{VP}(t))\right] + c_n^{VP}(r,t) + q^{II}(r,t) \tag{4}$$

$r$ is the slice position, $p_{\{1,2\}}(r)$ denote spatial basis functions[11] evaluated at position $r$, $d_{\{1,2\}}(t)$ and $d_{\{1,2\},n}^{VP}(t)$ are the corresponding field coefficients referring to the test gradient and the n$^{th}$ prephasing gradient, respectively. They are given explicitly in Equation (1). $c(r,t)$ and $c_n^{VP}(r,t)$ are phase contributions originating from concomitant fields, while $q^I(r,t)$ and $q^{II}(r,t)$ are background terms related to the slice selection gradient. By stacking the equations for the acquisitions from $M$ different slice positions, we obtain Equation (5). We solved this matrix equation for the maximum likelihood solution $d(t)$ with the method described by Harkins and Does.[4]

$$\begin{bmatrix} f_{1,1} \\ f_{2,1} \\ \vdots \\ f_{N,1} \\ f_{1,2} \\ f_{2,2} \\ \vdots \\ f_{N,2} \\ f_{1,3} \\ f_{2,3} \\ \vdots \\ f_{N,3} \\ f_{1,4} \\ f_{2,4} \\ \vdots \\ f_{N,4} \end{bmatrix} = \begin{bmatrix} P & P & 0 & \cdots & 0 & I_M & 0 & \cdots & 0 & 0 & 0 & \cdots & 0 \\ P & 0 & P & \cdots & 0 & 0 & I_M & \cdots & 0 & 0 & 0 & \cdots & 0 \\ \vdots & \vdots & \vdots & \ddots & \vdots & \vdots & \vdots & \ddots & \vdots & \vdots & \vdots & \ddots & \vdots \\ P & 0 & 0 & \cdots & P & 0 & 0 & \cdots & I_M & 0 & 0 & \cdots & 0 \\ -P & -P & 0 & \cdots & 0 & I_M & 0 & \cdots & 0 & 0 & 0 & \cdots & 0 \\ -P & 0 & -P & \cdots & 0 & 0 & I_M & \cdots & 0 & 0 & 0 & \cdots & 0 \\ \vdots & \vdots & \vdots & \ddots & \vdots & \vdots & \vdots & \ddots & \vdots & \vdots & \vdots & \ddots & \vdots \\ -P & 0 & 0 & \cdots & -P & 0 & 0 & \cdots & I_M & 0 & 0 & \cdots & 0 \\ 0 & P & 0 & \cdots & 0 & 0 & 0 & \cdots & 0 & I_M & 0 & \cdots & 0 \\ 0 & 0 & P & \cdots & 0 & 0 & 0 & \cdots & 0 & 0 & I_M & \cdots & 0 \\ \vdots & \vdots & \vdots & \ddots & \vdots & \vdots & \vdots & \ddots & \vdots & \vdots & \vdots & \ddots & \vdots \\ 0 & 0 & 0 & \cdots & P & 0 & 0 & \cdots & 0 & 0 & 0 & \cdots & I_M \\ 0 & -P & 0 & \cdots & 0 & 0 & 0 & \cdots & 0 & I_M & 0 & \cdots & 0 \\ 0 & 0 & -P & \cdots & 0 & 0 & 0 & \cdots & 0 & 0 & I_M & \cdots & 0 \\ \vdots & \vdots & \vdots & \ddots & \vdots & \vdots & \vdots & \ddots & \vdots & \vdots & \vdots & \ddots & \vdots \\ 0 & 0 & 0 & \cdots & -P & 0 & 0 & \cdots & 0 & 0 & 0 & \cdots & I_M \end{bmatrix} \begin{bmatrix} d \\ d_1^{VP} \\ d_2^{VP} \\ \vdots \\ d_N^{VP} \\ c_1^I \\ c_2^I \\ \vdots \\ c_N^I \\ c_1^{II} \\ c_2^{II} \\ \vdots \\ c_N^{II} \end{bmatrix} \quad (5)$$

$$\text{with} \quad f_{n,k} = \begin{bmatrix} f_{n,k}(r_1,t) \\ f_{n,k}(r_2,t) \\ \vdots \\ f_{n,k}(r_m,t) \\ \vdots \\ f_{n,k}(r_M,t) \end{bmatrix}, \quad P = \begin{bmatrix} 1 & r_1 \\ 1 & r_2 \\ \vdots & \vdots \\ 1 & r_m \\ \vdots & \vdots \\ 1 & r_M \end{bmatrix}, \quad d = \begin{bmatrix} \Delta B_0(t) \\ G(t) \end{bmatrix}, \quad d_n^{VP} = \begin{bmatrix} \Delta B_{0,n}^{VP}(t) \\ G_n^{VP}(t) \end{bmatrix},$$

$$c_n^I = \begin{bmatrix} c(r_1,t) + c_n^{VP}(r_1,t) + q^I(r_1,t) \\ c(r_2,t) + c_n^{VP}(r_2,t) + q^I(r_2,t) \\ \vdots \\ c(r_m,t) + c_n^{VP}(r_m,t) + q^I(r_m,t) \\ \vdots \\ c(r_M,t) + c_n^{VP}(r_M,t) + q^I(r_M,t) \end{bmatrix}, \quad c_n^{II} = \begin{bmatrix} c_n^{VP}(r_1,t) + q^{II}(r_1,t) \\ c_n^{VP}(r_2,t) + q^{II}(r_2,t) \\ \vdots \\ c_n^{VP}(r_m,t) + q^{II}(r_m,t) \\ \vdots \\ c_n^{VP}(r_M,t) + q^{II}(r_M,t) \end{bmatrix} \quad (6)$$

$n = 1, \ldots, N$ is the index of the prephasing step, $m = 1, \ldots, M$ is the slice number, $I_M$ is the $M \times M$ identity matrix, and $k = 1, 2, 3, 4$ iterates through the acquisition steps in Figure 1B.

In the case of CVP (Figure 1A), Equation (2) reduces to $f_{n,2}(r,t) = q^I(r,t)$ and Equation (4) reduces to $f_{n,4}(r,t) = q^{II}(r,t)$. To obtain a well-posed system of equations, we neglect the contributions from concomitant fields in this case, i.e. $c(r,t) = c_n^{VP}(r,t) = 0$. We thus arrive at the following matrix equation:

$$\begin{bmatrix} f_{1,1} \\ f_{2,1} \\ \vdots \\ f_{N,1} \\ f_{1,2} \\ f_{2,2} \\ \vdots \\ f_{N,2} \\ f_{1,3} \\ f_{2,3} \\ \vdots \\ f_{N,3} \\ f_{1,4} \\ f_{2,4} \\ \vdots \\ f_{N,4} \end{bmatrix} = \begin{bmatrix} P & P & 0 & \cdots & 0 & I_M & 0 \\ P & 0 & P & \cdots & 0 & I_M & 0 \\ \vdots & \vdots & \vdots & \ddots & \vdots & \vdots & \vdots \\ P & 0 & 0 & \cdots & P & I_M & 0 \\ 0 & 0 & 0 & \cdots & 0 & I_M & 0 \\ 0 & 0 & 0 & \cdots & 0 & I_M & 0 \\ \vdots & \vdots & \vdots & \ddots & \vdots & \vdots & \vdots \\ 0 & 0 & 0 & \cdots & 0 & I_M & 0 \\ 0 & P & 0 & \cdots & 0 & 0 & I_M \\ 0 & 0 & P & \cdots & 0 & 0 & I_M \\ \vdots & \vdots & \vdots & \ddots & \vdots & \vdots & \vdots \\ 0 & 0 & 0 & \cdots & P & 0 & I_M \\ 0 & 0 & 0 & \cdots & 0 & 0 & I_M \\ 0 & 0 & 0 & \cdots & 0 & 0 & I_M \\ \vdots & \vdots & \vdots & \ddots & \vdots & \vdots & \vdots \\ 0 & 0 & 0 & \cdots & 0 & 0 & I_M \end{bmatrix} \begin{bmatrix} d \\ d_1^{VP} \\ d_2^{VP} \\ \vdots \\ d_N^{VP} \\ q^I \\ q^{II} \end{bmatrix} \quad (7)$$

with $f_{n,k}$, $P$, $d$, $d_n^{VP}$ as in Equation (6), and $\quad q^I = \begin{bmatrix} q^I(r_1,t) \\ q^I(r_2,t) \\ \vdots \\ q^I(r_m,t) \\ \vdots \\ q^I(r_M,t) \end{bmatrix}, \quad q^{II} = \begin{bmatrix} q^{II}(r_1,t) \\ q^{II}(r_2,t) \\ \vdots \\ q^{II}(r_m,t) \\ \vdots \\ q^{II}(r_M,t) \end{bmatrix} \quad (8)$

The original "variable-prephasing" (VP) solution is derived from measurement steps (1) and (2) in Figure 1A. It can be found by solving the following matrix equation:

$$\begin{bmatrix} f_{1,1} \\ f_{2,1} \\ \vdots \\ f_{N,1} \\ f_{1,2} \\ f_{2,2} \\ \vdots \\ f_{N,2} \end{bmatrix} = \begin{bmatrix} P & I_M \\ P & I_M \\ \vdots & \vdots \\ P & I_M \\ 0 & I_M \\ 0 & I_M \\ \vdots & \vdots \\ 0 & I_M \end{bmatrix} \begin{bmatrix} d \\ q^I \end{bmatrix} \quad (9)$$

with $f_{n,k}$, $P$, $d$, and $q^I$ as in Equation (8) \hfill (10)

**Methods**

*Hardware*

Experiments were conducted on a 7T scanner (MAGNETOM Terra, Siemens Healthcare, Erlangen, Germany) equipped with a 32-channel head coil (Nova Medical, Wilmington, MA, USA). According to the manufacturer's specifications, maximum gradient amplitudes of 80 mT/m and maximum slew rates of 200 T/m/s were reachable on the scanner's gradient system. We did not disable the vendor's built-in eddy current compensation (ECC) for any of the measurements. A spherical phantom (165 mm diameter) filled with Polydimethylsiloxan oil was used for all experiments.

*Gradient waveform measurements*

First, we measured the waveform of a trapezoidal test gradient (amplitude 42 mT/m, ramp time 220 µs, flat top time 550 µs) with VP, CVP, and FCVP. We used the following sequence parameters: TR 500 ms, 9 slices, slice positions (distance from isocenter) 0 mm, ±4.0 mm, ±8.0 mm, ±12.5 mm, ±16.5 mm, slice thickness 2 mm, flip angle 90°, readout length 10.2 ms, dwell time 2.5 µs, 10 prephasing steps. Since finding the maximum likelihood solutions to the matrix equations (5), (7), and (9) involves the signal magnitudes as well as the signal phases, it was important to reach a steady-state magnetization for these measurements. The acquisitions in each slice were therefore preceded by 10 dummy excitations. The VP solution of the gradient waveform was obtained by only evaluating the readouts from measurement steps (1) and (2) in Figure 1A according to Equation (9). To increase robustness against noise, the measured frequency evolutions $f(t)$ were treated with a moving median filter of length 3 before solving the corresponding matrix equations.

Second, we measured the readout gradient train of an EPI readout. The parameters of the corresponding EPI acquisition are given in the next paragraph. The readout gradient reached maximum amplitudes of about 35 mT/m. For the waveform measurement, we used a readout length of 100 ms and a dwell time of 5 µs. The other parameters were identical to the measurements of the trapezoidal gradient waveform.

We evaluated the measured gradient waveforms by comparing the results from the VP, CVP, and FCVP measurements to one another, and by comparing them to a prediction based on the gradient system transfer function (GSTF).[11] The Fourier transform of the nominal gradient waveform was multiplied by the GSTF of the respective axis, and the result transformed back to the time domain. The nominal waveform in this calculation only consisted of the gradient of interest, i.e. the trapezoid or the EPI readout gradient, and did not contain the prephasing or slice selection gradients. Dwell time differences were accounted for by adding a delay correction to the GSTF prediction (-1 µs for the trapezoid, and -6.3 µs for the EPI readout gradient).

The GSTF was determined in the scope of a previous study.[10] Triangular probing gradients of different durations were measured with the thin-slice method with the same phantom and coil described above. These triangle measurements were combined in a linear system of equations, from which the GSTF was calculated by a regularized matrix inversion[18] and Fourier transform.

*Image acquisition and reconstruction*

We acquired a single-shot EPI image with the following parameters: TR 100 ms, TE 33 ms, FOV in readout direction 221 mm, FOV in phase encoding direction 80%, flip angle 90°, slice position isocenter, slice thickness 1.3 mm, matrix size 340 x 136, resolution 1.3 mm isotropic, partial Fourier factor 6/8, echo spacing 0.79 ms, receiver bandwidth 1470 Hz/pixel, ramp sampling on. We acquired a transversal slice with phase encoding in anterior-posterior direction.

We performed five different reconstructions:

1) For the first reconstruction, the phase difference between odd and even echoes was corrected based on the three navigator lines acquired prior to every EPI readout.[19] A constant phase shift, determined from the three reference echoes, was applied to the raw data of every second echo, and the nominal trajectory was used for the image reconstruction.

2) In the second reconstruction, the trajectory was calculated by applying the gradient system transfer function (GSTF). The GSTF-predicted waveform of the readout gradient was calculated as described above, and then integrated in the time domain.

3), 4), 5) In the third, fourth, and fifth reconstruction, the readout gradient trains measured with VP, CVP, and FCVP were used for the trajectory calculation. For the phase encoding direction, the nominal gradient waveform was assumed.

We used the non-uniform FFT (NUFFT) toolbox[20] from the Michigan Image Reconstruction Toolbox (MIRT)[21] for image reconstruction.

In all but the first reconstruction, small additional delay corrections compensating dwell time differences were required to minimize ghosting artifacts. They were implemented by delaying the corresponding gradient waveforms before integration to obtain the k-space trajectories. In order to optimize the delays, a circular mask covering the phantom was defined, and the image intensities outside of this circle were summed up for reconstructions with different delays. The minimum of this sum was found by fitting a polynomial of fourth degree to the results and determining the roots of the first derivative. The corresponding delay was regarded as "optimal" and applied in the respective reconstruction. The optimal delay determined for reconstructions 4 was also used in reconstructions 3 and 5.

The ghosting in the different reconstructions was quantitatively compared by means of the relative ghost intensity Γ, which we calculated as the ratio of the maxima of the signal intensities in two regions of interest (ROIs): One placed in a region with visible artifacts (ROI 1), and the other one shifted by half the FOV in the phase encoding direction (ROI 2).

## Results

### *Gradient waveforms*

Figure 2 displays the VP measurement of the trapezoidal test gradient on the x-axis and compares it to the GSTF prediction. The zoomed view in Figure 2B reveals a slight deviation between the two curves at around 0.6 ms. Panel C offers a closer look on the lingering field oscillations after the gradient is turned off, where the VP-measured gradient clearly differs from the GSTF prediction. Panel D shows the difference between measurement and prediction, which exhibits an oscillatory pattern. The start and end point of the trapezoid's ramps appear as spikes in the difference curve. In Panel E, the difference is overlaid with the measurement of the largest prephasing gradient in the respective time window, which was extracted from the CVP measurement. This curve matches well with the difference between the VP measurement and the GSTF prediction, except for the time points coinciding with the ramps of the trapezoid.

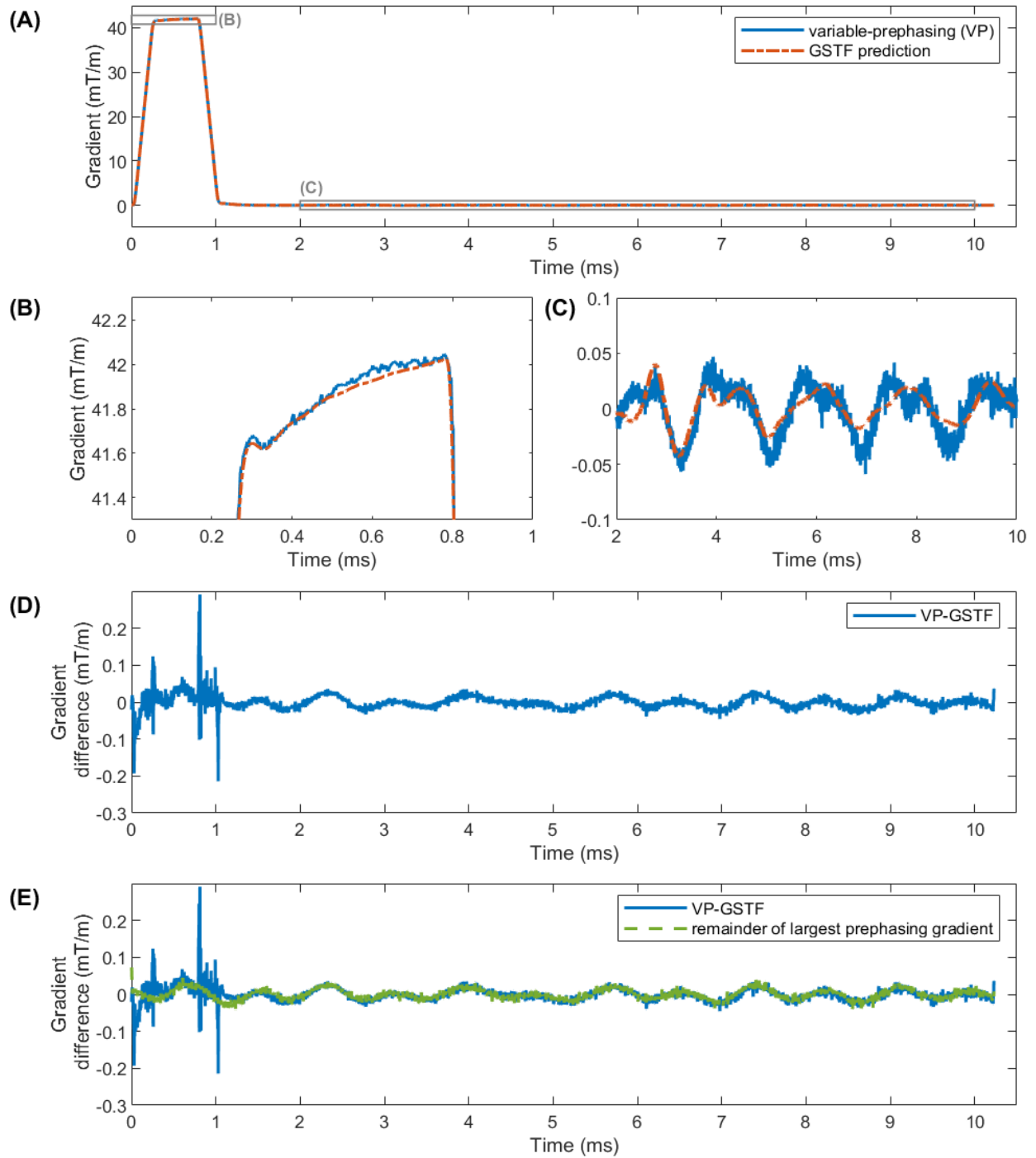

*Figure 2. Comparison of the measurement of the trapezoidal test gradient with variable-prephasing (VP) and the GSTF prediction. (A-C) Measured and predicted gradient waveforms. (D) Difference between VP measurement and GSTF prediction. The zoom in (C) and the difference in (D) reveal considerable discrepancies between the two gradient time courses in the lingering field oscillations after the test gradient is turned off. (E) Difference curve from (D) overlaid by the measured lingering oscillations of the largest prephasing gradient. For most of the measured time window, the two curves match well.*

Figure 3 compares the CVP and FCVP measurements of the trapezoidal test gradient to the GSTF prediction. Similar to the VP measurement, the CVP and FCVP measurements slightly deviate from the GSTF prediction at around 0.6 ms (cf. Figure 3B). The CVP measurement is slightly above and the FCVP measurement slightly below the GSTF-predicted waveform. In contrast to VP, both CVP and FCVP replicate the predicted lingering field oscillations shown in panel C almost identically. This is confirmed by the difference curves displayed in panel D. As already seen in Figure 2D, the switching points of the gradient waveform present as spikes in the difference.

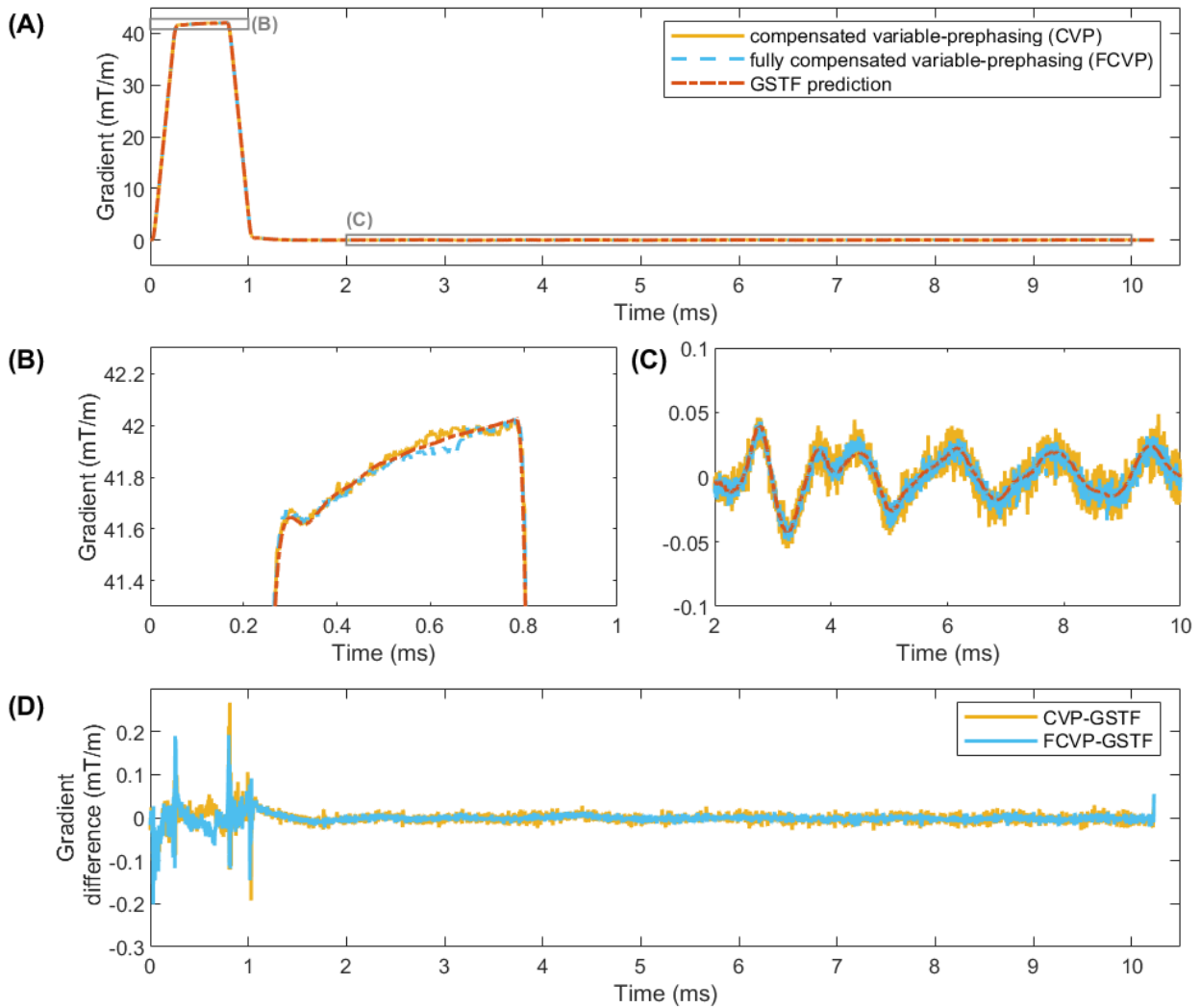

*Figure 3. Comparison of the measurement of the trapezoidal test gradient with compensated variable-prephasing (CVP), fully compensated variable-prephasing (FCVP), and the GSTF prediction. (A-C) Measured and predicted gradient waveforms. (D) Differences between the measurements and the GSTF prediction. The predicted lingering field oscillations of the test gradient are replicated almost identically by the two measurement methods.*

Figure 4A-D depict the VP, CVP, and FCVP measurements, and the GSTF prediction of the EPI readout gradient train on the x-axis. In panels C and D, we see that the CVP and the VP waveform agree well with each other and with the GSTF-predicted waveform. The FCVP waveform deviates visibly from these three for parts of the plateau phases of the gradient train (in panel C from about 11.5 ms onwards, and in panel D around 12.3 ms). It seems to be lower than the CVP and VP measurements for the second half of the positive lobes of the EPI gradient, and higher (i.e. less negative) than the CVP and VP measurements for the first half of the negative lobes. Figure 4E displays the difference between the VP and CVP

measurement of the EPI gradient. The difference exhibits an oscillatory pattern, similar to Figure 2D. In panels E and F, this difference is overlaid by a weighted sum of the lingering oscillations from the prephasing gradients, extracted from the CVP measurement. The weights for each time point mimic how the different prephasing steps contribute to the maximum likelihood solution of Equation (9), i.e. they are proportional to the quadratic signal magnitude of the respective prephasing step measured in box (1) in Figure 1A (averaged over the 9 slices). The zoomed view in panel G reveals that, overall, the two curves match well. Figure 4H and I show the difference between the CVP and FCVP measurement of the EPI gradient. It has a repetitive structure similar to a sawtooth wave. This confirms that the observations made in panel C and D extend over the whole gradient train.

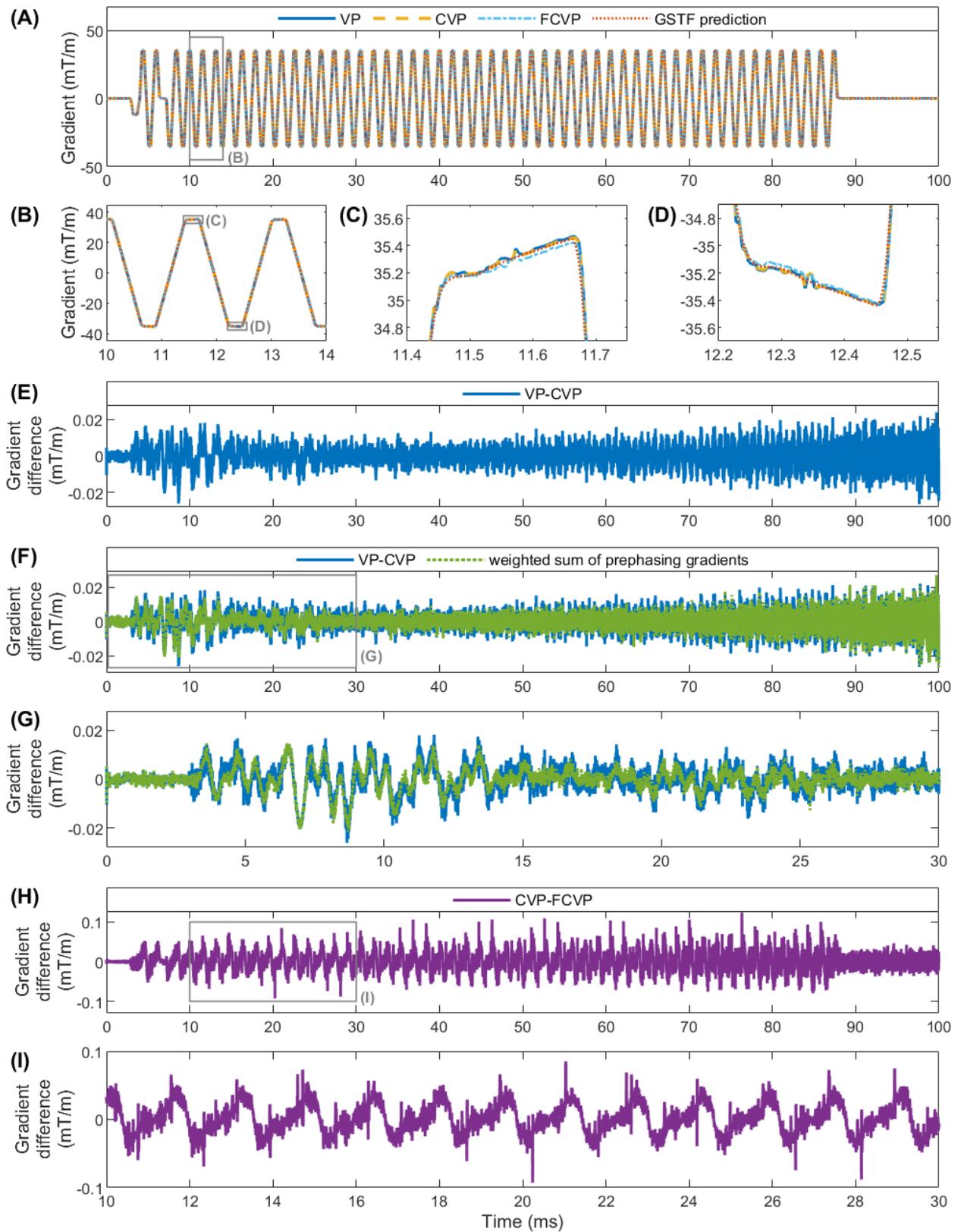

Figure 4. (A-D) EPI readout gradient waveform on the x-axis of the gradient system, measured with VP, CVP, and FCVP, and predicted by the GSTF. The FCVP measurement differs visibly from the other three. (E-G) Difference between the VP and CVP measurement. In (F) and (G), it is overlaid with a weighted sum of the lingering oscillations from the prephasing gradients, which agrees well with the difference curve. (H-I) Difference between the CVP and FCVP measurement. Its repetitive structure resembles a sawtooth wave.

*EPI images*

Figure 5 depicts the transverse EPI images in panels A-E. Panel A shows the image resulting from the standard ghost correction method based on three reference echoes acquired prior to the image readout. Nyquist ghosts are clearly visible. The image in panel B is reconstructed with the GSTF-based reconstruction applied to the readout- and phase encoding gradients (i.e. the physical x- and y-axis). An additional delay correction of -0.98 µs (as derived from the minimization of ghosting energy as described above) was applied in order to compensate dwell time differences between the GSTF measurement and the EPI acquisition. The ghosting artifacts are greatly reduced in this image, compared to panel A. Figure 5C-E present the reconstructions with the VP-, CVP-, and FCVP measurements of the readout gradient, respectively, in which the Nyquist ghosts are even further reduced. The additional delay correction compensating for the dwell time difference between gradient measurement and image acquisition was 7.28 µs in these reconstructions. All three images reconstructed with the measured gradient trains look similar. Figure 5F displays the k-space positions at the center point of each readout for the different reconstructions. The navigator-based reconstruction assumes the ideal case that $k_{RO} = 0$ at the center of each readout and corrects the phase of the measured MR signal accordingly (red dots). In the GSTF-based and measurement-based reconstructions, on the other hand, we leave the measured data as is and adjust the k-space trajectory. We see that in all these cases, the centers of the odd and even readouts are shifted against each other, and that the shift oscillates in the upper part of k-space, which is sampled first after the excitation. For the VP- and CVP-measured trajectories, there is an additional drift towards positive k-values.

We determined a relative ghost intensity of $\Gamma_{ref} = 26.1\%$ for the standard ghost correction method using reference echoes (Figure 5A). In the GSTF-based reconstruction (panel B), this value is almost halved to $\Gamma_{GSTF} = 13.4\%$. Both the reconstructions with the VP- (panel C) and CVP-measured readout gradient (panel D) exhibit the lowest relative ghost intensity of $\Gamma_{VP} = \Gamma_{CVP} = 9.2\%$. With the FCVP measurement of the readout gradient (panel E), the value is slightly higher again ($\Gamma_{FCVP} = 12.7\%$).

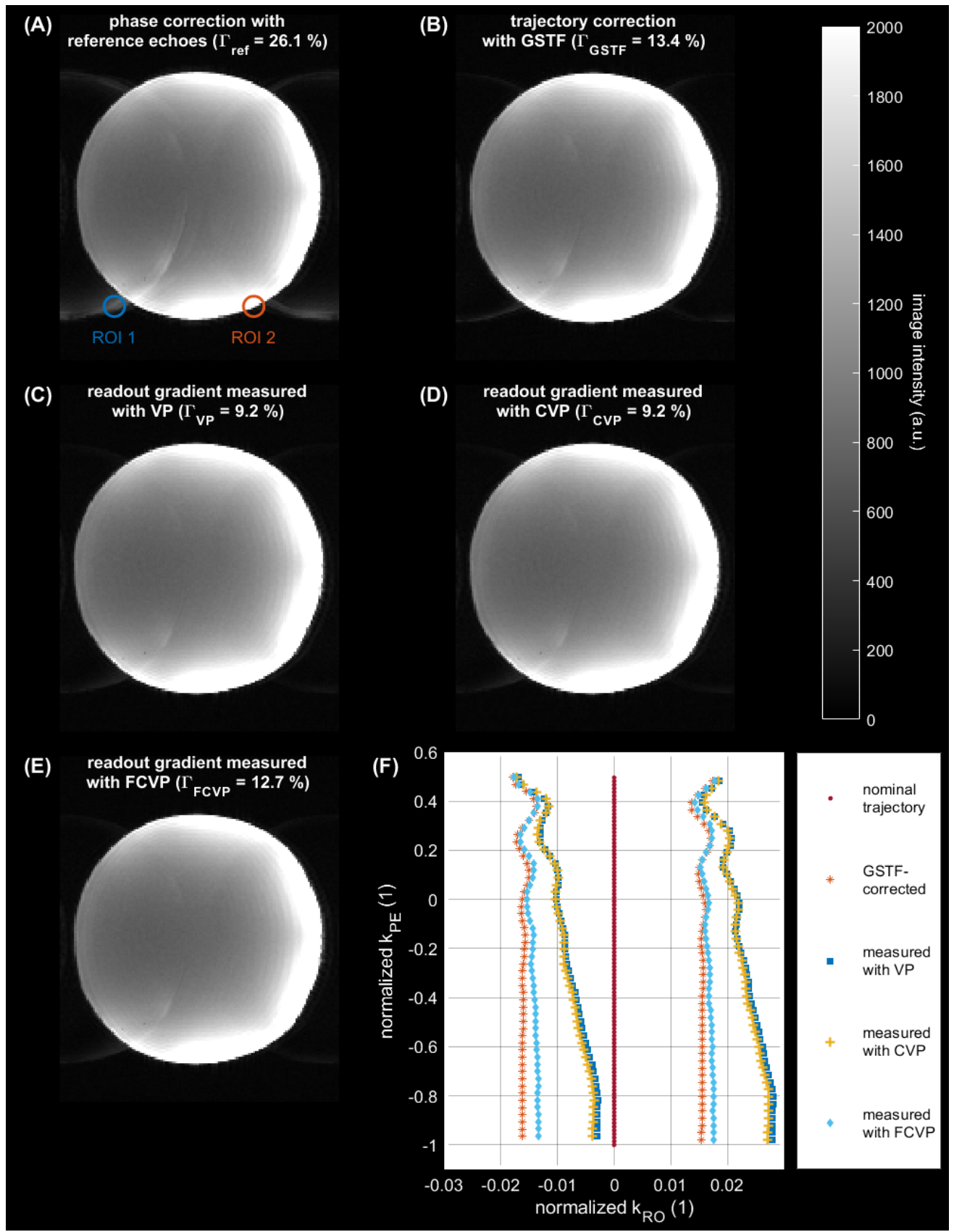

*Figure 5. Transverse EPI images, reconstructed with the (A) navigator-based, (B) GSTF-based, (C) VP measurement-based, (D) CVP measurement-based, and (E) FCVP measurement-based correction. Nyquist ghosts are most dominant in (A) and least visible in (C-E). The ROIs in (A) were used for ghost quantification by means of the relative ghost intensity $\Gamma$. (F) k-Space positions at the central points of each readout. In the GSTF-based and measurement-based trajectories, the shift between odd and even echoes clearly oscillates at the beginning of the EPI echo train.*

**Discussion and Conclusions**

We presented a new phantom-based measurement approach to determine the temporal evolution of magnetic gradient fields in an MRI scanner, namely "compensated variable-prephasing" (CVP), which is based on the "variable-prephasing" (VP) method[4]. "Fully compensated variable-prephasing" (FCVP) was already introduced in a previous study[10], but has now been examined more closely, and compared to VP and CVP. In both CVP and FCVP, we compensate lingering effects of the prephasing gradients, which are neglected in VP. Applying all three methods to a trajectory correction for an EPI image, we demonstrated that measuring the actual readout gradient yields superior suppression of ghosting artifacts compared to a navigator-based phase correction, or a GSTF-based trajectory correction. However, the differences we detected between the gradient waveforms measured with VP, CVP, and FCVP did not translate to visible differences in the EPI images for our setup at 7 T main field strength.

We observed lingering effects of the prephasing gradients when we compared the VP measurement of a trapezoidal test gradient to the GSTF prediction of the same waveform (cf. Figure 2). We noticed substantial differences in the lingering field oscillations after the gradient was switched off. These differences originated almost exclusively from the prephasing gradient with the largest absolute moment (cf. Figure 2E), which can be explained by the way Equation (9) is solved. To find the maximum likelihood solution to the matrix equation, essentially, the phase signals measured with the different prephasing gradients are weighted by the squared corresponding signal magnitude for each time point. Details on this can be found in the publication of the VP method[4], and our provided MATLAB (MathWorks, Natick, MA, USA) code. Since the largest prephasing gradient rephases the signal at the end of the test gradient, the signal magnitude is largest after the test gradient has been switched off. Therefore, the VP measurement superimposes the lingering oscillations of this prephasing gradient with the lingering oscillations of the trapezoid itself. During the test gradient, on the other hand, the signals from multiple prephasing steps contribute to the final result. The first prephasing steps have higher weights in the beginning of the test gradient, the middle ones dominate in the center of the test gradient, and the last ones are most relevant for the time points at the end of the test gradient.

In CVP and FCVP, the lingering oscillations of each prephasing gradient are measured separately and can thus be distinguished from the signal of the test gradient. As a result, both measurements agree extremely well with the GSTF prediction in the time window after the trapezoidal test gradient (cf. Figure 3). Only when the gradient is active, we still see significant deviations between the measurements and the prediction (cf. Figure 3B, D). We attribute these differences to nonlinearities in the gradient amplifier behavior[22–25], which can have an impact on our observations in two ways:

1) The nonlinearities are not replicated by the GSTF model, causing inaccuracies in the gradient prediction. We believe that this is the dominant cause for the deviations during the ramps and at the switching points of the trapezoid (cf. Figure 3D).
2) In both FCVP and the GSTF measurement, we use reference measurements in which the gradient of interest is inverted, to compensate for concomitant field effects. However, this compensation only works accurately when the actual output gradient with inverted sign is the exact negative of the non-inverted output gradient. Nonlinearities in the gradient amplifiers can violate this assumption. We think that this is the main reason for the deviations between CVP measurement, FCVP measurement, and GSTF prediction during the plateau of the trapezoidal test gradient (cf. Figure 3B).

Since the feedback loop in the gradient amplifiers will always introduce certain nonlinearities into the gradient signal chain, we conclude that the CVP method yields the most accurate phantom-based measurement of gradient waveforms at high and ultrahigh field strengths. The acquisition time for CVP could theoretically be further reduced by 25% by skipping step (2) in Figure 1A and lengthening the ADC in step (4), such that the measurement window from step (2) is also covered in step (4). Concomitant fields of the lowest order are inversely proportional to $B_0$[17], and can thus be expected to be negligible above 1 T. At low (e.g. 0.55 T) or ultralow field MRI scanners, however, FCVP might be the superior method, given that concomitant field contributions might supersede the nonlinearity effects. Further research is required to separate both effects. Looking at measurements of the gradient amplifiers' output currents could be a promising approach.[22–24]

Our measurements of an EPI readout gradient in Figure 4 demonstrate that the uncompensated lingering oscillations of the prephasing gradients in the VP method can also alter the measurement result during active gradient switching (cf. Figure 4E-G). How much they contribute at each time point depends again on the respective signal magnitude. The systematic differences we found between the CVP and FCVP measurements of the EPI gradient (cf. Figure 4H, I) confirm that the symmetry of positive and negative gradients, which we assume in FCVP, is violated in our measurements. If the difference came from uncompensated concomitant fields in the CVP measurement, the deviation would be in the same direction for positive and negative lobes of the EPI gradient, since concomitant fields of the lowest order depend quadratically on the gradient amplitude[17]. However, we find that the sign of the difference changes from the positive to the negative gradient lobes.

The high level of detail with which we detected differences between the presented measurement methods testifies to the high precision of the methods. However, it tells us little about how relevant this precision is for actual imaging applications. We chose EPI as an application for demonstration, because gradient inaccuracies cause Nyquist ghosting (or N/2 ghosting) in EPI images[26], which is an easily identifiable artifact. Due to the alternating readout directions in EPI, errors alternate as well between odd and even k-space lines. As a result, the signal in k-space is modulated with half the Nyquist frequency, and a copy of the object, shifted by half the field of view (FOV), appears as a ghost in the image. The current standard method to correct for N/2 ghosts, which is implemented on most clinical MRI scanners, uses two or three reference echoes to determine the shift between odd and even echoes, which is then compensated for.[19] It is assumed that this shift is constant for all phase encoding lines, which is reasonably fulfilled on clinical scanners with 1.5 T or 3 T magnets. However, it has been shown before that on human size 7 T systems, the shift between odd and even echoes varies substantially throughout the entire EPI readout.[27,28] The N/2 ghosts are therefore not adequately corrected by the standard method. Figure 5A confirms this, and Figure 5F demonstrates how exactly the shift varies in terms of which k-space position is reached at the center of each readout. For the GSTF predicted trajectory, and the FCVP measurement, we observe a damped oscillation, as it has been described before.[27,28] This oscillation occurs because the main frequency of the EPI readout gradient is very close to a mechanical resonance on the same gradient axis (in this case the x-axis). These mechanical vibrations of the gradient coils are stronger at 7 T than at lower field strengths[16,25], or certain systems with a different architecture, for example small-animal scanners.[29] The distance of the main peak of the EPI spectrum from mechanical resonances depends on the chosen sequence parameters. The closer it gets, the stronger are the occurring oscillations. Therefore, precise knowledge of the actual k-space trajectory can be a helpful asset to remove ghosting from EPI images on human 7 T scanners.

In the VP and CVP measurements of the trajectory, the centrally sampled k-space positions additionally drift in positive $k_{RO}$ direction. In principle, a sheared k-space trajectory results in a shear of the imaged object.[30] In our experiments, this effect is negligibly small though. While there are no directly visible differences in ghosting intensity between the three measurement-based reconstructions (Figure 5C-E), our quantitative evaluation revealed that the VP- and CVP measurements of the readout gradient work best for ghost suppression (with a relative ghosting intensity of 9.2 %). The reconstruction with the FCVP measurement exhibits a slightly higher relative ghosting intensity of 12.7 %, which is similar to the relative ghosting intensity of the GSTF-based reconstruction (13.4 %). This correlates with the amount of drifting we see in the respective central readout points. Drifting could be related to the nonlinearities occurring during gradient ramping (cf. Figure 2D and Figure 3D), which are present in the EPI gradient train in a similar manner as in the trapezoidal test gradient (data not shown). However, a more detailed analysis of this matter is beyond the scope of this study. In summary, Figure 5 shows that gradient measurements with VP and CVP are equivalent as far as their application in EPI image reconstruction at our 7 T scanner is concerned. Using a GSTF predicted trajectory for image reconstruction yields slightly inferior results in terms of ghost suppression but is still considerably better than the navigator-based standard method, as it halves the measured relative ghost intensity. The quantitative results emphasize that the CVP method's relative insensitivity to gradient amplifier nonlinearities directly translates into its potential for artifact reduction. This could justify its superiority over the other methods in possible clinical setups at high or ultrahigh field.

While we only analyzed first-order gradient self-terms in this work, the proposed CVP method – as well as FCVP and the original VP approach – can be extended to also determine higher-order terms of the gradient field dynamics. This can be achieved by adding phase encoding gradients to the sequence, similar to the approach for higher-order GSTF measurements published by Rahmer et al.[31] In the mathematical formulation of our measurement schemes in Equations (1) - (4), the sum over the spatial basis functions $p_j(r)$ simply has to be extended to higher indices, and the submatrix $P$ (c.f. Equation (5)) has to be complemented by the appropriate basis functions.[11]

In conclusion, both CVP and FCVP expressed differences in the measured gradient waveforms compared to VP and the GSTF predictions. Since the FCVP approach is potentially compromised by a sign asymmetry in the gradient signal chain, we consider CVP the most reliable gradient measurement technique for our experimental setup. It enables a phantom-based determination of the gradient field evolution with high precision, which can be useful for trajectory corrections in non-Cartesian or single-shot imaging techniques. We demonstrated its application in single-shot EPI imaging, where it proved to yield superior ghost suppression compared to the standard navigator-based correction approach, a GSTF-based trajectory correction, or correcting the trajectory with the FCVP measurement of the readout gradient.

**Conflicts of interest**

The University Hospital Würzburg has a research collaboration agreement with Siemens Healthineers AG.

**Data Availability Statement**

The measurement data and MATLAB code used to generate the results in this work are publicly available at https://zenodo.org/doi/10.5281/zenodo.13742003.